\documentclass[12pt]{article}

\usepackage{epsfig}
\usepackage{amsmath}
\usepackage{amssymb}
\usepackage{axodraw}

\providecommand{\beqa}{\begin{eqnarray}}
\providecommand{\eeqa}{\end{eqnarray}}

\providecommand{\Dta}{{\overline{D3}}}
\providecommand{\Mta}{{\overline{M2}}}
\providecommand{\Mfa}{{\overline{M5}}}

\def\cD{{\cal{D}}}
\def\cH{{\cal{H}}}
\def\cM{{\cal{M}}}
\def\cO{{\cal{O}}}
\def\cS{{\cal{S}}}

\begin{document}

\begin{titlepage}
\begin{flushright}
UMD-PP-04-023\\
hep-th/0312311\\
\today
\end{flushright}

\vspace{1cm}
\begin{center}
\baselineskip25pt {\Large\bf Black Holes, Space-Filling Chains and
Random Walks}

\end{center}
\vspace{1cm}
\begin{center}
\baselineskip12pt {Axel Krause\footnote{E-mail: now at Jefferson Physical Laboratory, Harvard University, Cambridge, MA 02138, USA}}
\vspace{1cm}

{\it Center for String and Particle Theory,}\\[1.8mm]
{\it Department of Physics, University of Maryland}\\[1.8mm]
{\it College Park, MD 20742, USA}

\vspace{0.3cm}
\end{center}
\vspace*{\fill}

\begin{abstract}
Many approaches to a semiclassical description of gravity lead to
an integer black hole entropy. In four dimensions this implies
that the Schwarzschild radius obeys a formula which describes the
distance covered by a Brownian random walk. For the
higher-dimensional Schwarzschild-Tangherlini black hole, its
radius relates similarly to a fractional Brownian walk. We propose
a possible microscopic explanation for these random walk
structures based on microscopic chains which fill the interior of
the black hole.
\end{abstract}

\noindent
PACS: 05.20.-y, 05.40.Jc, 11.25.Uv, 97.60.Lf\\
Keywords: Black Holes, Random Walks, D-Branes

\vspace*{\fill}

\end{titlepage}

\section{Introduction}

The full non-perturbative formulation of quantum gravity continues
to pose a formidable challenge. Though loop quantum gravity had
some success in dealing with 4d non-supersymmetric black holes, we
believe, following string-theory, that to incorporate the
electroweak and strong gauge interactions next to gravity we have
to leave four dimensions and start with ten or eleven dimensions
in the first place. The appearance of four non-compact dimensions
must then be explained by some dynamical partial
decompactification process enlarging four of the ten resp.~eleven
dimensions during the cosmic evolution of the very early universe.
The success of this idea hinges very much on our ability to better
understand M-theory at a non-perturbative level. Progress in this
direction has been made by the proposal of M-theory as a
matrix-theory \cite{Matrix}.

Also on the frontier of deriving predictions from M/string-theory
in order to address phenomenology and test the theory, decisive
developments took place over the last few years. Various moduli
have been stabilized by means of fluxes, non-perturbative effects
or a combination of both (see e.g.~\cite{ModStab1}-\cite{ModStab10}). To really see
whether M/string-theory is a phenomenologically viable theory, it
seems most promising to consider heterotic M-theory \cite{HW1},\cite{HW2} and compactify it down to four dimensions \cite{HetFlux1}-\cite{HetFlux5}. Though the
effective supergravity description allows a phenomenologically
realistic stabilization of most moduli, one realizes that
ultimately the structure of the complete theory requires the
knowledge of the full non-perturbative quantum M-theory. This
problem was already pointed out in the original work \cite{HW1},\cite{HW2}.

Given therefore the urgent fundamental and phenomenological need
to understand the microscopic formulation of M/string-theory, one
of the best guidance principles could be the Bekenstein-Hawking
(BH) entropy \cite{BHE1}-\cite{BHE7}, which every candidate quantum gravity theory should successfully explain microscopically. In this
respect string-theory was remarkably successful in explaining the
entropy of certain supersymmetric extremal and near-extremal black
holes correctly - without having to adjust an arbitrary
proportionality constant appropriately \cite{StringBH1}-\cite{StringBH3}. What is
however still problematic in conventional string-theory is the
counting of the microscopic degrees of freedom directly in the
non-perturbative regime where the weakly coupled string technology
is no longer applicable. This, however, is required when the black
hole is non-supersymmetric and for instance of Schwarzschild or
Kerr type.

By starting from a 10 or 11-dimensional spacetime whose event
horizon and complete internal compactification space are wrapped
by two Euclidean electric-magnetic dual brane pairs
$(E_1,M_1)+(E_2,M_2)$, progress towards explaining the BH-entropy
directly in the non-perturbative regime was made recently in
\cite{K11}-\cite{K17}. Namely by interpreting the inverse tension
of a Euclidean brane as a smallest volume unit on the brane which
is a direct consequence of the `worldvolume uncertainty principle'
\cite{CHK}, the wrapped Euclidean branes can be thought of as
being composed out of a finite number of smallest cells (units of
smallest worldvolume). On this lattice-like structure chain
excitations were proposed \cite{K11},\cite{K17} whose
microcanonical ensemble entropy correctly reproduces both the
BH-entropy and its known leading logarithmic correction. For some
other earlier interesting ideas on how to understand the entropy
of non-supersymmetric black holes from string-theory see
\cite{BHS1}-\cite{BHS9}.

\section{Black Holes and Brownian Random Walks}

One of the results of the proposal made in \cite{K11,K17} has been
that the BH-entropy equals an integer
\beqa
\cS_{BH} = N \in \mathbf{N} \; ,
\label{BHInt}
\eeqa
which represents the total number of cells on the dual brane pair
lattice. This relation arose not only in four \cite{K11} but also
in higher dimensions \cite{K17}. Moreover, it appeared in many
other approaches to black hole quantization, as well \cite{BHN1}-\cite{BHN3}.
Let us study its consequences for the $d$-dimensional
hyperspherically symmetric Schwarzschild-Tangherlini black hole
with topology $\mathbf{R}\times\mathbf{S}^{d-2}$. These black
holes are described by a metric \cite{MP}
\beqa
ds^2 = - \left(1-\left(\frac{r_S}{r}\right)^{d-3}\right) dt^2
+ \left(1-\left(\frac{r_S}{r}\right)^{d-3}\right)^{-1} dr^2
+ r^2 d\Omega^2_{d-2}
\eeqa
with $r$ the radial coordinate and $d\Omega^2_{d-2}$ the
line-element on the unit $d-2$ sphere. We will consider values
$d\ge 4$ (in $d=3$ a vanishing Ricci-tensor implies a flat
spacetime and conventional black holes are absent).

Given the mass $M_d$ of the black hole, its Schwarzschild radius
$r_S$ is defined as
\beqa
M_d = \frac{(d-2)A_{d-2}}{16 \pi G_d} r_S^{d-3} \; ,
\qquad A_{d-2} = \frac{2\pi^{\frac{(d-1)}{2}}}{\Gamma\!\left(\frac{d-1}{2}\right)}
\label{Mass}
\eeqa
with $A_{d-2}$ the `area' of the unit $d-2$ sphere. The BH-entropy
for these black holes is given by the standard expression
\beqa
\cS_{BH} = \frac{vol(\cH_{d-2})}{4G_d}
\label{BHE}
\eeqa
where $G_d$ is the $d$-dimensional Newton's Constant and
$\cH_{d-2}$ the black hole surface of topology $\mathbf{S}^{d-2}$
with `area'
\beqa
vol(\cH_{d-2}) = r_S^{d-2}A_{d-2} \; .
\label{Area}
\eeqa

A direct consequence of the integer-valued BH-entropy
(\ref{BHInt}) in conjunction with (\ref{BHE}) and (\ref{Area}) is
that the Schwarzschild radius $r_S$ becomes a discrete function of
the integer $N$
\beqa
r_S(N) =  l_d \left(\frac{4}{A_{d-2}}\right)^{\frac{1}{d-2}}
N^{\frac{1}{d-2}} \; ,
\label{dWalk}
\eeqa
where the $d$-dimensional Planck-length $l_d$ is given by
\beqa
l_d^{d-2} = G_d \; .
\eeqa
In four dimensions, for $d=4$, this relation becomes
\beqa
r_S(N) = \frac{l_4}{\sqrt{\pi}} \sqrt{N} \; ,
\label{4Walk}
\eeqa
which is precisely the relation for the root-mean-square (rms)
distance covered by a {\em Brownian random walk} of average
step-size $l_4/\sqrt{\pi}$. In general, a fractional Brownian
random walk of step-size $l_{walk}$ and dimension $d_{walk}$ is
characterized by the relation (see e.g.~\cite{FBW})
\beqa
r_{walk} = l_{walk} N^{\frac{1}{d_{walk}}}
\label{FRW}
\eeqa
expressing the rms-distance covered, $r_{walk}$, as a function of
the number of steps $N$. Hence, we recognize that as a consequence
of the integer-valued BH-entropy the Schwarzschild radius $r_S$ of
a $d$-dimensional black hole is given by a {\em fractional
Brownian random walk} relation with step-size
\beqa
l_{walk} = l_d \left(\frac{4}{A_{d-2}}\right)^{\frac{1}{d-2}}
\label{FracStep}
\eeqa
and integer dimension
\beqa
d_w = d-2 \ge 2 \; .
\eeqa
Though a fractional Brownian random walk is mathematically also
defined for fractional dimensions $d_w\notin\mathbf{N}$ not being
integers, this possibility is not realized by black holes where
all $d_{walk}$ happen to be integers.

As an aside, let us remark that a fractional Brownian walk can
also be regarded as a diffusion process with `time' $\tau
= N\Delta\tau$ and `time-increment' $\Delta\tau$ (during $\Delta\tau$
the diffusion spreads out over the step-length $l_{walk}$). The
diffusion coefficient $\cD$ is defined via the diffusion equation
\beqa
r_w^2 = 2 \cD \tau \; .
\eeqa
Hence, with (\ref{FRW}) the diffusion coefficient becomes
\beqa
\cD =
\left(\frac{l_{walk}}{\sqrt{2}\Delta\tau^{1/d_{walk}}}\right)^2
\frac{1}{\tau^{(d_{walk}-2)/d_{walk}}}
=
\frac{C(d_{walk})}{\tau^{(d_{walk}-2)/d_{walk}}}
\; ,
\eeqa
with some constant $C(d_{walk})$ depending on $d_{walk}$. One learns that the higher-dimensional $d > 4$ case with $d_{walk}
> 2$ corresponds to an anomalous diffusion coefficient, i.e.~one
with non-trivial `time' $\tau$ dependence, while the 4-dimensional
$d=4$ case with $d_{walk} = 2$ is distinguished by a truly
constant $\tau$ independent diffusion coefficient. Thus for the
higher-dimensional case $\cD$ decreases with `time' $\tau$ and it
becomes increasingly difficult for the diffusion process to spread
out as `time' $\tau$ progresses. Hence, in this case the diffusion
process is called `subdiffusion' \cite{JFG}. Such processes are
known to occur as well in disordered or poorly connected media in
which the subdiffusion process can be studied e.g.~by the `ant in
a labyrinth' model of de Gennes describing a particle constrained
to diffuse through a percolation cluster \cite{AIL}. It would be
interesting to understand better this distinction of four
dimensions which we will, however, not attempt in this work.

An interesting and important property of a fractional Brownian
random walk of dimension $d_{walk}$ is its capability to fill the
volume of a finite space of dimension $d_{walk}$ completely in the
large $N$ limit \cite{FBW}, \cite{JFG}. This generalizes the
well-known result of a Brownian random walk which covers a
2-dimensional space in the large $N$ limit. The pressing question
which arises is whether and how these random walk behaviors
occurring for black holes can be understood directly from the
microscopic physics leading to the integer-valuedness of the
BH-entropy (\ref{BHInt}) in the first place. We will attempt in
the next section a simple explanation within the chain-approach
developed in \cite{K11}-\cite{K17}. This approach has its origin
in non-perturbative M/string-theory and therefore requires us to
start in ten resp.~eleven dimensions. But before going into
details here, let us mention yet another consequence of
(\ref{BHInt}). Namely, by using the random walk results for the
Schwarzschild radius and plugging it into the mass formula
(\ref{Mass}), the $d$-dimensional black hole's mass $M_d$ becomes
a discrete function of $N$ as well
\beqa
M_d(N) = \frac{(d-2)}{4\pi}
\left(\frac{A_{d-2}}{4}\right)^{\frac{1}{d-2}}
\times\frac{N^{\frac{d-3}{d-2}}}{l_d} \; .
\eeqa
This expression becomes more suggestive when written in terms of
the Schwarzschild radius as
\beqa
M_d(N) = \frac{(d-2)}{4\pi}\times\frac{N}{r_S(N)} \; ,
\eeqa
in which form it resembles the $N$th Kaluza-Klein excitation of a
field compactified on a circle of radius $4\pi r_S/(d-2)$. It
might therefore be an interesting hint at the microscopic quantum
gravitational Hamiltonian.

\section{Space-Filling Chains}

The approach of \cite{K11}-\cite{K17} to understand the BH-entropy
in terms of microscopic chain-states was proposed to address the
problem of black hole state counting directly in the
non-perturbative regime where the black hole lives. Though firmly
rooted in M/string-theory, this microscopic derivation of the
BH-entropy uses a direct counting mechanism which does not rely on
supersymmetry and is therefore also applicable to
non-supersymmetric spacetimes such as Schwarzschild black holes or
their higher-dimensional cousins, the Schwarzschild-Tangherlini
black holes. Let us be a bit more specific about this approach.

Starting with 10-dimensional type II string-theory (alternatively
one could start from 11-dimensional M-theory \cite{K11,K17}), a
$d$-dimensional spacetime $\cM^{1,d-1}$ with $d-2$ dimensional
horizon $\cH^{d-2}$ (the intersection of the spacetime's future
event horizon with a partial Cauchy surface) is embedded into ten
dimensions through the direct product $\cM^{1,d-1} \times
\cM^{10-d}$, with $\cM^{10-d}$ being compact. This describes a
compactification from ten to $d$ dimensions. For the uncharged,
non-dilatonic Schwarzschild-Tangherlini black holes one would
consider a doublet of Euclidean $(D3,D3)+(\Dta,\Dta)$ brane pairs
\cite{K1213},\cite{K17}, each wrapped around $\cH^{d-2} \times
\cM^{10-d}$ where $\cH^{d-2} \simeq \mathbf{S}^{d-2}$. Here, the
first entry in each brane pair is always orthogonal to the second.
By applying the `brane worldvolume uncertainty principle'
\cite{CHK} which splits the brane worldvolume into a discrete
number of smallest cells, whose size is given by the inverse of
the brane's tension \cite{K11}, one concludes that the joined
worldvolume of the $(D3,D3)+(\Dta,\Dta)$ doublet consists out of a
finite number of $N$ cells. The presence of the branes allows for
a simple rewriting of the $d$-dimensional BH-entropy associated
with the spacetime $\cM^{d-1}$, implying the entropy's
integer-valuedness (\ref{BHInt}).

With $N$ the number of cells on the $(D3,D3)+(\Dta,\Dta)$
worldvolume, the natural excitations to look at (in the
non-perturbative regime where the string coupling constant $g_s$
is of $\cO(1)$ and the finite cell sizes, being proportional to
$g_s^2$, become relevant) are links starting and ending on any of
these cells. Indeed chains composed out of $N$ such links (closed
chain case) or $N-1$ links (open chain case) possess a
microcanonical ensemble entropy which correctly reproduces the
$d$-dimensional BH-entropy and its logarithmic correction
\cite{K11,K17}. It is these chains which exhibit already an
inherent random structure which will be of interest to us now.
Actually such discrete random-walk structures are also suggested
by string-theory in the weakly coupled regime, namely at high
temperatures near the Hagedorn temperature. Here, it was known for
a long time (see e.g.~\cite{MT1},\cite{MT2}) that the string in this regime is best described by a random walk. Moreover, composites of open strings which could generate chain structures in string theory appeared in \cite{CC}.

Our aim will now be to show that the fractional Brownian random
walk relation (\ref{dWalk}) can arise from such a chain when it is
assumed to fill the interior of the black hole, bounded by the
horizon $\cH^{d-2}\simeq \mathbf{S}^{d-2}$, i.e.~the $d-1$
dimensional hyperball $\mathbf{D}^{d-1}$. To this end we will need
to determine the volume covered by the chain.

\subsection{D=10 Type II String-Theory Case}

\setcounter{figure}{0}
\begin{figure}[t]
\begin{center}
\begin{picture}(260,120)(0,20)
\CArc(130,80)(60,0,360)
\DashCurve{(76,78)(85,73)(92,70)}{2}
\DashCurve{(78,68)(86,63)(93,60)}{2}
\DashCurve{(79,79)(80,70)(82,62)}{2}
\DashCurve{(88,75)(89,66)(91.5,59)}{2}
\DashCurve{(162,105)(171,101)(178,96)}{2}
\DashCurve{(164,95)(173,91)(179,86)}{2}
\DashCurve{(165,107)(167,97)(167.5,89)}{2}
\DashCurve{(174,103)(176,92)(176.5,85)}{2}
\Line(79.5,75.8)(166.5,103.3)
\Line(88.5,71.5)(175.3,98.2)
\Line(81,66)(167.5,94)
\Line(90,61)(176,88.5)
\LongArrow(130,22)(130,78)
\LongArrow(130,78)(130,22)
\Text(139,50)[]{$r_S$}
\LongArrow(60,80)(84,70)
\Text(49,83)[]{$V_{cell}$}
\LongArrow(110,105)(127,88)
\Text(99,111)[]{$\text{link}$}
\LongArrow(91,57)(177,84.5)
\LongArrow(177,84.5)(91,57)
\Text(152,64)[]{$L_{link}$}
\LongArrow(210,120)(182.5,113)
\Text(225,122)[]{$\mathbf{S}^{d-2}$}
\LongArrow(205,55)(167,50)
\Text(221,57)[]{$\mathbf{D}^{d-1}$}
\end{picture}
\caption{Each link starts and ends on any of the cells of the
brane complex. Since (part of) the branes wrap the
(hyper-)spherical black hole horizon of radius $r_S$, most links
stretch right through the interior of the (hyper-)sphere and thus
have length $L_{link}\approx r_S$.}
\label{Links}
\end{center}
\end{figure}
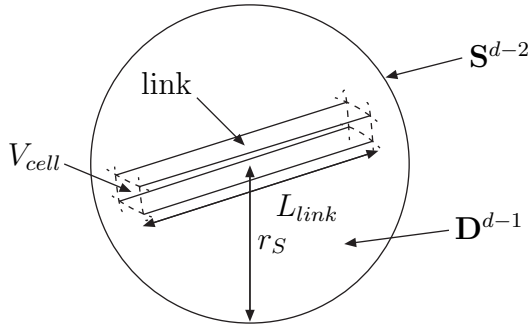
For the purpose of explaining the $d$-dimensional fractional
random walk relation (\ref{dWalk}), it will turn out to be
sufficient to assume that the size of the internal
compactification manifold, $\cM^{10-d}$, is much smaller than the
Schwarzschild radius $r_S$ of the external $d$-dimensional black
hole in $\cM^{1,d-1}$. This means that on the internal
$\cM^{10-d}$ all links will end on just a very few cells while
there is much more freedom for the links in $\cM^{1,d-1}$ to end
on many different cells on the much larger horizon sphere
$\mathbf{S}^{d-2}$. Hence, it will be enough to consider only the
freedom of the links to vary over the external horizon sphere. As,
by assumption, each link has the freedom to start and end on any
of the cells covering $\mathbf{S}^{d-2}$ with equal probability,
most links will then stretch right through the interior of the
hyperball $\mathbf{D}^{d-1}$ which is bounded by the horizon
sphere $\mathbf{S}^{d-2}$ (see fig.\ref{Links}).

\begin{figure}[ht]
  \begin{center}
  \epsfig{file=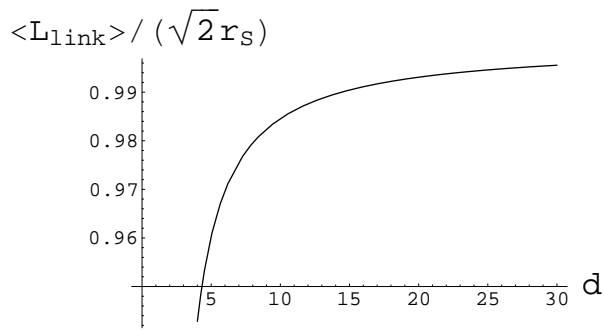,width=8cm,angle=0}
  \caption{The convergence of the average link length
          $\langle L_{link}\rangle$ towards $\sqrt{2}r_S$ for large
          values of the dimension $d$.}
\label{LinkConvergence}
  \end{center}
\end{figure}

The average length of a single link $\langle L_{link} \rangle$ is
determined by keeping one end of the link fixed and letting the
other vary over all $N$ cells
\beqa
\langle L_{link} \rangle
= \frac{1}{N}\sum_{i=1}^{N} L_{link,i} \; ,
\eeqa
where $L_{link,i}$ is the value $L_{link}$ assumes when the link
stretches between the fixed cell and the $i$th cell. In view of
the assumed smallness of the internal compactification radius, the
length $L_{link,i}$ reduces to the length the link stretches
inside $\mathbf{D}^{d-1}$. The cell volume, being given by
\cite{K17}
\beqa
V_{cell} = 8 l_d^{d-2} vol(\cM^{10-d})
\label{VCell}
\eeqa
implies that the cell size lies between the $d$-dimensional
Planck-length and the compactification radius. Hence, it should be
considerably smaller than $r_S$ as well, and we are entitled to
approximate the average over the discrete cells by an integral
average over the hypersphere $\mathbf{S}^{d-2}$
\begin{alignat}{3}
\langle L_{link}\rangle &=
\frac{1}{A_{d-2}}\int_{\mathbf{S}^{d-2}} \!\!\!\!\! L_{link}
\;d\Omega_{d-2}
\label{LinkAve1} \\
&= \frac{1}{A_{d-2}} \int_0^{2\pi} d\phi \int_0^\pi\sin\theta_1
d\theta_1 \int_0^\pi\sin^2\theta_2 d\theta_2 \hdots
\int_0^\pi\sin^{d-3}\theta_{d-3} d\theta_{d-3} \;L_{link} \; .
\label{LinkAve2}
\end{alignat}
Here $(\theta_{d-3},\hdots,\theta_2,\theta_1,\phi)$ are spherical
coordinates parameterizing $\mathbf{S}^{d-2}$. The result which
will be derived in the appendix is
\begin{table}[t]
\begin{center}
\begin{tabular}{|c||c|c|c|c|c|c|c|}
\hline $d$ & 4 & 5 & 6 & 7 & 8 & 9 & 10 \\
\hline $\langle L_{link}\rangle/r_S$ & $1.333$ & $1.358$ & $1.371$
& $1.380$ & $1.385$ & $1.389$ & $1.392$ \\
\hline
\end{tabular}
\caption{The dependence of the average link length $\langle
L_{link}\rangle$ in units of the Schwarzschild radius $r_S$ on the
dimension $d$ of the external spacetime.}
\label{LinkTable}
\end{center}
\end{table}
\beqa
\langle L_{link}\rangle = 2r_S
\frac{B(\frac{d-1}{2},d-2)}{B(\frac{d}{2}-1,d-\frac{3}{2})}
= \frac{2^{d-2}\,\Gamma^2\!\left(\frac{d-1}{2}\right)}
{\sqrt{\pi}\,\Gamma\!\left(d-\frac{3}{2}\right)} \, r_S \; ,
\label{LinkAve3}
\eeqa
where $B(x,y)$ denotes the Euler Beta-function. It leads to the
numerical values displayed in table \ref{LinkTable}. The value for
$\langle L_{link}\rangle$ increases slowly but monotonically with
$d$. By using Stirling's formula for the asymptotic behavior of
the Gamma-function one can show that $\langle L_{link}\rangle$
converges for large $d$ towards
\beqa
\langle L_{link}\rangle \rightarrow \sqrt{2} r_S \; , \quad d\gg 1
\; .
\eeqa
This behavior is shown in fig.\ref{LinkConvergence}. The
asymptotic value $\sqrt{2} r_S$ corresponds to a link which
stretches e.g.~from the north pole to the equator. In critical
M/string-theory (for the M-theory case see next subsection 3.2),
of course, only the values $d \le 11$ will be relevant.

Since the size of the 8-dimensional cross-section of each link is
always given by $V_{cell}$, it follows that the average volume
occupied by a link is
\beqa
V_{link} = \langle L_{link} \rangle V_{cell} \; .
\eeqa
The average volume of the chain, $V_{chain}$, is then given by
multiplication with $N$, the number of links which the chain
contains (as we will work in the large $N$ limit, there is no need
to distinguish between `closed' chains with $N$ links and `open'
chains with $N-1$ links, see \cite{K17} -- both will occupy the
same volume)
\beqa
V_{chain} = N V_{link} = N \langle L_{link} \rangle V_{cell} \; .
\label{VChain}
\eeqa

The next step will consist in considering chains at large $N$, the
regime where one connects to black holes of macroscopic size.
Because of the random nature of the chains -- each link can start
and end on any of the $N$ cells with the same probability -- such
chains tend to cover at sufficiently large $N$ the full finite
space $D^{d-1}\times\cM^{10-d}$ through which they stretch. Notice
that the filling of the $\mathbf{D}^{d-1} \times \cM^{10-d}$ space
by the chains does not occur due to a fractal behavior of the
chain, as the links already possess the same dimensionality, nine,
as the space $\mathbf{D}^{d-1} \times \cM^{10-d}$. For such large
$N$ space-filling chains one is led to the following volume
equality
\beqa
1 = \frac{\text{Volume of Chain}}{\text{Volume of }
\mathbf{D}^{d-1}\times\cM^{10-d}}
= \frac{N \langle L_{link} \rangle V_{cell}}
{A_{d-2} \big(\frac{r_S^{d-1}}{d-1}\big) vol(\cM^{10-d})}
\label{SpaceFill}
\eeqa
By plugging in the expression (\ref{VCell}) for $V_{cell}$, the
internal volume drops out of this equation. Employing furthermore
the result (\ref{LinkAve3}) for $\langle L_{link} \rangle$, we
obtain
\beqa
\frac{\text{Volume of Chain}}{\text{Volume of }
\mathbf{D}^{d-1}\times\cM^{10-d}}
= \left(\frac{l_d}{r_S}\right)^{d-2} N f(d)
\label{VolRatio}
\eeqa
where the function
\beqa
f(d) = (d-1)\frac{2^d\, \Gamma^3\!\left(\frac{d-1}{2}\right)}
{\pi^{d/2}\, \Gamma\!\left(d-\frac{3}{2}\right)}
\eeqa
depends only on the dimension $d$. With this relation the equality
(\ref{SpaceFill}) leads to the following relation for the radius
$r_S(N)$ of the ball $\mathbf{D}^{d-1}$
\beqa
r_S(N) = l_d f(d)^{\frac{1}{d-2}} N^{\frac{1}{d-2}} \; .
\label{dChainWalk}
\eeqa

\begin{figure}[t]
  \begin{center}
  \epsfig{file=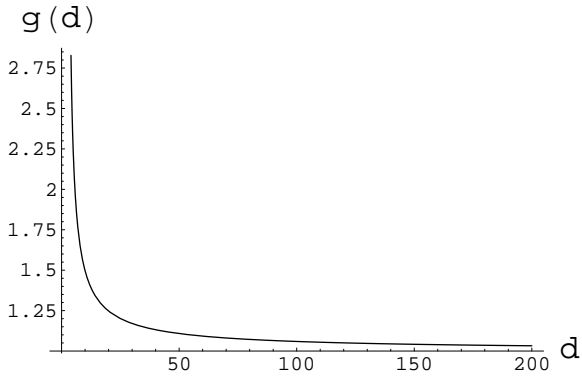,width=8cm,angle=0}
  \caption{The convergence of the ratio of the step-sizes,
           $g(d)$, towards $1$ for large values of the dimension $d$.}
\label{RatioConvergence}
  \end{center}
\end{figure}

Since the hypersurface $\mathbf{S}^{d-2}$ has been identified with
the horizon boundary $\cH^{d-2}$ of the Schwarzschild-Tangherlini
black hole, we see that the chains at large $N$ which become
$\mathbf{D}^{d-1} \times \cM^{10-d}$ space-filling, give indeed
the correct fractional Brownian random walk relation $r_S(N)
\propto l_d N^{1/(d-2)}$ in each dimension $d$. Though the
step-sizes of the walks (\ref{dWalk}) and (\ref{dChainWalk}) do
not agree exactly, their difference is not large and even vanishes
when $d$ becomes large. To see this quantitatively, let us look at
the ratio of both step-sizes given by
\beqa
g(d) = \left(\frac{f(d)}{4/A_{d-2}}\right)^{\frac{1}{d-2}} \; .
\eeqa
It approaches unity for large values of $d$ (see
fig.\ref{RatioConvergence})
\beqa
g(d) \simeq \sqrt{2}^{\frac{1}{d-2}} \rightarrow 1 \; ,
\quad d\gg 1
\label{equality}
\eeqa
as one can see by using Stirling's approximation for the
Gamma-function at large argument. For the physical values of $d$
ranging from 4 to 10 in type II string-theory we collect the
numerical values of $g(d)$ in table \ref{RatioTable}.
\begin{table}[b]
\begin{center}
\begin{tabular}{|c||c|c|c|c|c|c|c|}
\hline $d$ & 4 & 5 & 6 & 7 & 8 & 9 & 10 \\
\hline $g(d)$ & $2.83$ & $2.22$ & $1.92$
& $1.75$ & $1.64$ & $1.56$ & $1.50$ \\
\hline
\end{tabular}
\caption{The ratio of the step-sizes, $g(d)$, as a function of the
dimension $d$.}
\label{RatioTable}
\end{center}
\end{table}

Let us comment on the possible origin and resolution of this small
discrepancy between the step-sizes at finite $d$. One observes
that $g(d)$ is always larger than one. This means that the chain
volume as given by (\ref{VChain}) is slightly bigger than required
in order to give the correct step-size (\ref{FracStep}). Notice
that it is only the $d$-dependent numerical factor contained in
(\ref{VChain}) which is slightly too large -- the dependence on
all the other parameters $N, r_S, l_d, vol(\cM^{10-d})$ is the
correct one. The fact that both step-sizes become equal in the
large $d$ limit already hints at a possible resolution. For this,
let us note that when a chain crosses in a random way a finite
space so often that eventually it fills this space completely,
self-intersections of the chain become inevitable. Notice further
that the chains are not infinitely thin but have a non-zero
8-dimensional transverse cross-section given by $V_{cell}$. The
chains can therefore intersect also in higher dimensions $d>4$ in
contrast to infinitely thin 1-dimensional mathematical chains
which cannot intersect in dimensions bigger than two (or if
behaving as Brownian walks with fractional dimension 2 such
mathematical chains wouldn't intersect in dimensions bigger than
four). Moreover, when self-intersections occur, one would expect
that the probability for them would decrease the larger the
dimension $d$ becomes. This simply because for higher $d$ there is
more `room' for two randomly placed links to avoid each other.
Therefore, by including also the effect of chain
self-intersections, one would expect to reduce the $d$-dependent
factor in the chain volume while maintaining the large $d$
equality (\ref{equality}). This would then hopefully reconcile
both step-sizes with each other. We will, however, not pursue this
effect further here.

\subsection{D=11 M-Theory Case}

Let us finally address the 11-dimensional M-theory case which
proceeds in an analogous manner. Here, the neutral
Schwarzschild-Tangherlini black holes can only be associated with
the charge-neutral combination of Euclidean $(M2,M5)+(\Mta,\Mfa)$
brane pairs wrapping $\cH^{d-2} \times \cM^{11-d}$, with
$\cM^{11-d}$ the compact internal space. The different, now $11-d$
dimensional, internal compactification space will result in a
different cell volume \cite{K17}
\beqa
V_{cell} = 8l_d^{d-2}vol(\cM^{11-d}) \; .
\label{MVCell}
\eeqa
Apart from this, the analysis in the D=11 case will be exactly the
same as in D=10. In particular, under the same assumption that
$r_S$ is much larger than the size of the compact $\cM^{11-d}$,
the average (\ref{LinkAve3}) of the link-length over the
hypersphere $\mathbf{S}^{d-2}\simeq \cH^{d-2}$ will be unaffected.

The chains will now be $\mathbf{D}^{d-1} \times \cM^{11-d}$ space
filling at large $N$ which leads to the equality
\beqa
1 = \frac{\text{Volume of Chain}}{\text{Volume of }
D^{d-1}\times\cM^{11-d}}
= \frac{N \langle L_{link} \rangle V_{cell}}
{A_{d-2} \big(\frac{r_S^{d-1}}{d-1}\big) vol(\cM^{11-d})} \; .
\label{MSpaceFill}
\eeqa
Since the different internal volume $vol(\cM^{11-d})$ drops out by
virtue of (\ref{MVCell}), this ratio coincides exactly with the
one for the D=10 case, given in (\ref{VolRatio}). Consequently one
obtains also for the D=11 M-theory case the same fractional
Brownian walk relation (\ref{dChainWalk}) as in the D=10 analysis
and thus a potential microscopic explanation of the semiclassical
random walk relation (\ref{dWalk}).

\bigskip
\noindent {\large \bf Acknowledgements}\\[2ex]
This work has been supported by the National Science Foundation
under Grant Number PHY-0099544.

\begin{appendix}

\section{The Average Link Length $\langle L_{link}\rangle$}

\begin{figure}[ht]
\begin{center}
\begin{picture}(260,120)(0,20)
\CArc(130,80)(60,0,360)
\LongArrow(210,120)(182.5,113)
\Text(223,122)[]{$\mathbf{S}^{d-2}$}
\Text(130,148)[]{$NP$}
\LongArrow(131,138.5)(187,67)
\Text(172,108)[]{${\vec L}_{link}$}
\LongArrow(130,80)(130,138.5)
\Text(114,105)[]{${\vec r}_{NP}$}
\Text(141,103)[]{$r_S$}
\LongArrow(130,79)(186.5,65.5)
\Text(157,62)[]{$\Delta{\vec L}_{link}$}
\end{picture}
\caption{In the regime where the cell size is much smaller than
the radius $r_S$ of the hypersphere $\mathbf{S}^{d-2}$, the links
become approximately thin lines and their average length over the
hypersphere can be determined by integration. The picture defines
the vectors used in the calculation.}
\label{Sphere}
\end{center}
\end{figure}
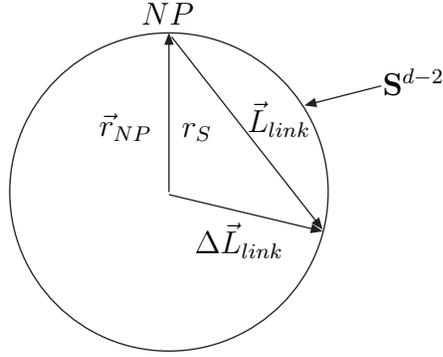

We will provide here the calculation of the average link-length
$\langle L_{link} \rangle$, deriving the result (\ref{LinkAve3}).
The definition of $\langle L_{link}\rangle$ was given in
(\ref{LinkAve2}). Let us first introduce standard spherical
coordinates
\begin{alignat}{3}
x_{d-1} &= r_s\cos\theta_{d-3} \notag \\
x_{d-2} &= r_s\sin\theta_{d-3}\cos\theta_{d-4} \notag \\
\vdots \\
x_2 &= r_s\sin\theta_{d-3}\hdots\sin\theta_1\cos\phi \notag \\
x_1 &= r_s\sin\theta_{d-3}\hdots\sin\theta_1\sin\phi \notag
\end{alignat}
which parameterize the hypersphere $\mathbf{S}^{d-2}$ defined by
the relation
\beqa
x_1^2+\hdots x_{d-1}^2 = r_S^2 \; .
\label{DefSphere}
\eeqa
With the Cartesian coordinate origin at the center of the sphere,
the north pole $NP$ (see fig.\ref{Sphere}) is described by
\beqa
{\vec r}_{NP} = (x_1=0,\hdots,x_{d-2}=0,x_{d-1}=r_S) \; .
\eeqa
Without loss of generality, we can fix for the determination of
the average link length the starting point of all links at the
north pole while the end-point of the links, described by the
vector ${\vec L}_{link}$ (see fig.\ref{Sphere}) varies over
$\mathbf{S}^{d-2}$. In terms of the difference vector (see
fig.\ref{Sphere})
\beqa
\Delta {\vec L}_{link} = (x_1,\hdots,x_{d-1})
\eeqa
${\vec L}_{link}$ becomes
\beqa
{\vec L}_{link} = \Delta {\vec L}_{link} - {\vec r}_{NP}
= (x_1,\hdots,x_{d-2},x_{d-1}-r_S) \; ,
\eeqa
where the Cartesian coordinates obey (\ref{DefSphere}).
$L_{link}$, the norm of the vector ${\vec L}_{link}$, can then be
expressed in spherical coordinates as
\beqa
L_{link} = 2r_S\sin\left(\frac{\theta_{d-3}}{2}\right) \; .
\eeqa

The integration (\ref{LinkAve2}) can now be performed trivially
over $\phi,\theta_1,\hdots,\theta_{d-4}$ and contributes the
`area' $A_{d-3}$ of the unit $d-3$ sphere
\beqa
\langle L_{link}\rangle = 2r_S\frac{A_{d-3}}{A_{d-2}}I_{d-3} \; .
\eeqa
where
\beqa
I_{d-3} = \int_0^\pi d\theta_{d-3} \sin^{d-3}\theta_{d-3}
\sin\left(\frac{\theta_{d-3}}{2}\right) \; .
\eeqa
It remains to determine this integral which can be done by
switching to the integration variable $x = \theta_{d-3}/2$
\beqa
I_{d-3} = 2^{d-2}\int_0^{\pi/2} dx \sin^{d-2} x \cos^{d-3} x \; .
\eeqa
This integral is equal to (see e.g.~\cite{AS})
\beqa
I_{d-3}
= 2^{d-3}\frac{\Gamma\big(\frac{d-1}{2}\big)
\Gamma\big(\frac{d-2}{2}\big)}{\Gamma\big(d-\frac{3}{2}\big)}
\eeqa
which can be brought into the simpler form
\beqa
I_{d-3} = \sqrt{\pi}\frac{\Gamma(d-2)}
{\Gamma\big(d-\frac{3}{2}\big)}
\eeqa
by using Legendre's duplication formula for the product of two
Gamma-functions.

Writing the `areas' $A_{d-2}$ and $A_{d-3}$ explicitly out using
(\ref{Mass}) one arrives at
\beqa
\langle L_{link}\rangle = 2r_S
\frac{\Gamma\big(\frac{d-1}{2}\big)\Gamma(d-2)}
{\Gamma\big(\frac{d}{2}-1\big)\Gamma\big(d-\frac{3}{2}\big)}
= 2r_S \frac{B\big(\frac{d-1}{2},d-2\big)}
{B\big(\frac{d}{2}-1,d-\frac{3}{2}\big)}
\eeqa
where the Beta-function allows a quite compact expression.
However, more useful for us will be an expression which is
obtained by employing once more Legendre's duplication formula in
the form
\beqa
\Gamma\Big(\frac{d}{2}-1\Big)
= \frac{\sqrt{\pi}\,\Gamma(d-2)}
{2^{d-3}\,\Gamma\big(\frac{d-1}{2}\big)}
\eeqa
to finally obtain the result (\ref{LinkAve3}) quoted in the main
text
\beqa
\langle L_{link}\rangle = r_S
\frac{2^{d-2}\,\Gamma^2\!\big(\frac{d-1}{2}\big)}
{\sqrt{\pi}\,\Gamma\big(d-\frac{3}{2}\big)} \; .
\eeqa

\end{appendix}

\newcommand{\zpc}[3]{{\sl Z.Phys.} {\bf C\,#1} (#2) #3}
\newcommand{\npb}[3]{{\sl Nucl.Phys.} {\bf B\,#1} (#2) #3}
\newcommand{\npps}[3]{{\sl Nucl.Phys.Proc.Suppl.} {\bf #1} (#2) #3}
\newcommand{\plb}[3]{{\sl Phys.Lett.} {\bf B\,#1} (#2) #3}
\newcommand{\prd}[3]{{\sl Phys.Rev.} {\bf D\,#1} (#2) #3}
\newcommand{\prb}[3]{{\sl Phys.Rev.} {\bf B\,#1} (#2) #3}
\newcommand{\pr}[3]{{\sl Phys.Rev.} {\bf #1} (#2) #3}
\newcommand{\prl}[3]{{\sl Phys.Rev.Lett.} {\bf #1} (#2) #3}
\newcommand{\prsla}[3]{{\sl Proc.Roy.Soc.Lond.} {\bf A\,#1} (#2) #3}
\newcommand{\jhep}[3]{{\sl JHEP} {\bf #1} (#2) #3}
\newcommand{\cqg}[3]{{\sl Class.Quant.Grav.} {\bf #1} (#2) #3}
\newcommand{\prep}[3]{{\sl Phys.Rep.} {\bf #1} (#2) #3}
\newcommand{\fp}[3]{{\sl Fortschr.Phys.} {\bf #1} (#2) #3}
\newcommand{\nc}[3]{{\sl Nuovo Cimento} {\bf #1} (#2) #3}
\newcommand{\nca}[3]{{\sl Nuovo Cimento} {\bf A\,#1} (#2) #3}
\newcommand{\lnc}[3]{{\sl Lett.~Nuovo Cimento} {\bf #1} (#2) #3}
\newcommand{\ijmpa}[3]{{\sl Int.J.Mod.Phys.} {\bf A\,#1} (#2) #3}
\newcommand{\rmp}[3]{{\sl Rev. Mod. Phys.} {\bf #1} (#2) #3}
\newcommand{\ptp}[3]{{\sl Prog.Theor.Phys.} {\bf #1} (#2) #3}
\newcommand{\sjnp}[3]{{\sl Sov.J.Nucl.Phys.} {\bf #1} (#2) #3}
\newcommand{\sjpn}[3]{{\sl Sov.J.Particles\& Nuclei} {\bf #1} (#2) #3}
\newcommand{\splir}[3]{{\sl Sov.Phys.Leb.Inst.Rep.} {\bf #1} (#2) #3}
\newcommand{\tmf}[3]{{\sl Teor.Mat.Fiz.} {\bf #1} (#2) #3}
\newcommand{\jcp}[3]{{\sl J.Comp.Phys.} {\bf #1} (#2) #3}
\newcommand{\cpc}[3]{{\sl Comp.Phys.Commun.} {\bf #1} (#2) #3}
\newcommand{\mpla}[3]{{\sl Mod.Phys.Lett.} {\bf A\,#1} (#2) #3}
\newcommand{\cmp}[3]{{\sl Comm.Math.Phys.} {\bf #1} (#2) #3}
\newcommand{\jmp}[3]{{\sl J.Math.Phys.} {\bf #1} (#2) #3}
\newcommand{\pa}[3]{{\sl Physica} {\bf A\,#1} (#2) #3}
\newcommand{\nim}[3]{{\sl Nucl.Instr.Meth.} {\bf #1} (#2) #3}
\newcommand{\el}[3]{{\sl Europhysics Letters} {\bf #1} (#2) #3}
\newcommand{\aop}[3]{{\sl Ann.~of Phys.} {\bf #1} (#2) #3}
\newcommand{\jetp}[3]{{\sl JETP} {\bf #1} (#2) #3}
\newcommand{\jetpl}[3]{{\sl JETP Lett.} {\bf #1} (#2) #3}
\newcommand{\acpp}[3]{{\sl Acta Physica Polonica} {\bf #1} (#2) #3}
\newcommand{\sci}[3]{{\sl Science} {\bf #1} (#2) #3}
\newcommand{\nat}[3]{{\sl Nature} {\bf #1} (#2) #3}
\newcommand{\pram}[3]{{\sl Pramana} {\bf #1} (#2) #3}
\newcommand{\desy}[1]{{\sl DESY-Report~}{#1}}
\newcommand{\hepth}[1]{{\tt hep-th/}{\tt #1}}
\newcommand{\hepph}[1]{{\tt hep-ph/}{\tt #1}}
\newcommand{\grqc}[1]{{\tt gr-qc/}{\tt #1}}
\newcommand{\astroph}[1]{{\tt astro-ph/}{\tt #1}}

\bibliographystyle{plain}

\begin{thebibliography}{99}

\bibitem{Matrix} T.~Banks, W.~Fischler, S.H.~Shenker and
L.~Susskind, \prd{55}{1997}{5112}, \hepth{9610043}.

\bibitem{ModStab1} S.~Gukov, C.~Vafa and E.~Witten, \npb{584}{2000}{69}, \hepth{9906070}.

\bibitem{ModStab2} S.~Giddings, S.~Kachru and J.~Polchinski, \prd{66}{2002}{106006}, \hepth{0105097}.

\bibitem{ModStab3} G.~Curio and A.~Krause, \npb{643}{2002}{131}, \hepth{0108220}.

\bibitem{ModStab4} S.~Kachru, M.~Schulz and S.~Trivedi, \jhep{0310}{2003}{007}, \hepth{0201028}.

\bibitem{ModStab5} S.~Kachru, R.~Kallosh, A.~Linde and S.P.~Trivedi, \prd{68}{2003}{046005}, \hepth{0301240}.

\bibitem{ModStab6} K.~Becker, M.~Becker, K.~Dasgupta, P.S.~Green, \jhep{0304}{2003}{007}, \hepth{0301161}.

\bibitem{ModStab7} G.L.~Cardoso, G.~Curio, G.~Dall'Agata and  D.~L\"ust, \jhep{0310}{2003}{004}, \hepth{0306088}.

\bibitem{ModStab8} E.I.~Buchbinder and B.A.~Ovrut, \prd{69}{2004}{086010}, \hepth{0310112}.

\bibitem{ModStab9} S.~Gukov, S.~Kachru, X.~Liu and L.~McAllister, \prd{69}{2004}{086008}, \hepth{0310159}.

\bibitem{ModStab10} M.~Becker, G.~Curio and A.~Krause, \npb{693}{2004}{223}, \hepth{0403027}.

\bibitem{HW1} P.~Ho\v{r}ava and E.~Witten, \npb{460}{1996}{506},
\hepth{9510209}.

\bibitem{HW2} P.~Ho\v{r}ava and E.~Witten, \npb{475}{1996}{94}, \hepth{9603142}.

\bibitem{HetFlux1} E.~Witten, \npb{471}{1996}{135},
\hepth{9602070}.

\bibitem{HetFlux2} A.~Lukas, B.A.~Ovrut and D.~Waldram, \npb{532}{1998}{43},\hepth{9710208}.

\bibitem{HetFlux3} A.~Lukas, B.A.~Ovrut and D.~Waldram,  \npb{540}{1999}{230}, \hepth{9801087}.

\bibitem{HetFlux4} G.~Curio and A.~Krause, \npb{602}{2001}{172}, \hepth{0012152}.

\bibitem{HetFlux5} G.~Curio and A.~Krause, \npb{693}{2004}{195}, \hepth{0308202}.

\bibitem{BHE1} D.~Christodoulou, \prl{25}{1970}{1596}.

\bibitem{BHE2} S.W.~Hawking, \prl{26}{1971}{1344}.

\bibitem{BHE3} J.M.~Bardeen, B.~Carter and
S.W.~Hawking, \cmp{31}{1973}{161}.

\bibitem{BHE4} J.~Bekenstein, \prd{7}{1973}{2333}.

\bibitem{BHE5} J.~Bekenstein, \prd{9}{1974}{3292}.

\bibitem{BHE6} S.W.~Hawking, \nat{248}{1974}{30}.

\bibitem{BHE7} S.W.~Hawking, \cmp{43}{1975}{199}.

\bibitem{StringBH1} A.~Strominger and C.~Vafa, \plb{379}{1996}{99},
\hepth{9601029}.

\bibitem{StringBH2} C.~Callan and J.~Maldacena,
\npb{472}{1996}{591}, \hepth{9602043}.

\bibitem{StringBH3} G.~Horowitz and A.~Strominger, \prl{77}{1996}{2368}, \hepth{9602051}.

\bibitem{K11} A.~Krause, \ijmpa{20}{2005}{4055},
\hepth{0201260}.

\bibitem{K1213} A.~Krause, \ijmpa{20}{2005}{2813},
\hepth{0205310}.

\bibitem{K1415} A.~Krause, \cqg{20}{2003}{S533}.

\bibitem{K17} A.~Krause, to appear in {\sl Int.J.Mod.Phys.} {\bf A}, \hepth{0312309}.

\bibitem{BHS1} L.~Susskind and J.~Uglum, \prd{50}{1994}{2700},
\hepth{9401070}.

\bibitem{BHS2} L.~Susskind and J.~Uglum, \npps{45BC}{1996}{115}, \hepth{9511227}.

\bibitem{BHS3} K.~Hotta, \ptp{99}{1998}{427}, \hepth{9705100}.

\bibitem{BHS4} T.~Banks, W.~Fischler, I.R.~Klebanov and L.~Susskind, \prl{80}{1998}{226}, \hepth{9709091}.

\bibitem{BHS5} I.R.~Klebanov and L.~Susskind, \plb{416}{1998}{62}, \hepth{9709108}.

\bibitem{BHS6} M.~Li, \jhep{9801}{1998}{009}, \hepth{9710226}.

\bibitem{BHS7} T.~Banks, W.~Fischler, I.R.~Klebanov and L.~Susskind, \jhep{9801}{1998}{008}, \hepth{9711005}.

\bibitem{BHS8} N.~Ohta and J.G.~Zhou, \npb{522}{1998}{125},
\hepth{9801023}.

\bibitem{BHS9} S.~Das, A.~Ghosh and P.~Mitra,
\prd{63}{2001}{024023}, \hepth{0005108}.

\bibitem{BHN1} J.D.~Bekenstein, \lnc{11}{1974}{467}.

\bibitem{BHN2} V.~Mukhanov, \jetpl{44}{1986}{63}.

\bibitem{BHN3} J.D.~Bekenstein and V.~Mukhanov, \plb{360}{1995}{7}, \grqc{9505012}; and references
contained in the second paper of ref.~[8].

\bibitem{MP} R.C.~Myers and M.J.~Perry, \aop{172}{1986}{304}.

\bibitem{FBW} B.B.~Mandelbrot, {\it The Fractal Geometry of
Nature}, W.H.~Freeman and Company (1983).

\bibitem{JFG} J.F.~Gouyet, {\it Physics and Fractal Structures},
Springer-Verlag (1996).

\bibitem{AIL} P.G.~de Gennes, {\sl La Recherche} {\bf 72} (1976) 919.

\bibitem{CHK} C.S.~Chu, P.M.~Ho and Y.C.~Kao,
\prd{60}{1999}{126003}, \hepth{9904133}.

\bibitem{MT1} D.~Mitchell and N.~Turok,
\prl{58}{1987}{1577}.

\bibitem{MT2} D.~Mitchell and N.~Turok, \npb{294}{1987}{1138}.

\bibitem{CC} A.~Krause, \npb{748}{2006}{98}, \hepth{0006226}.

\bibitem{AS} M.~Abramowitz and I.A.~Stegun, {\it Handbook of
Mathematical Functions}, Dover Publications (1965).

\end{thebibliography}

\end{document}